\documentstyle[prd,aps]{revtex}

\begin{document}

\newtheorem{prop}{Proposition}[subsection]
\renewcommand{\topfraction}{1.0}
\twocolumn[\hsize\textwidth\columnwidth\hsize\csname
@twocolumnfalse\endcsname

\title{Exact inhomogeneous Einstein-Maxwell-Dilaton cosmologies}

\author{ Stoytcho S.~Yazadjiev }

\address{Department of Theoretical  Physics, Faculty of Physics,
Sofia University,\\ 5 James Bourchier Boulevard, Sofia 1164,
Bulgaria
\\ {\tt E-mail: yazad@phys.uni-sofia.bg}}


\maketitle

\begin{abstract}
We present solution generating techniques which permit to
construct exact inhomogeneous and anisotropic cosmological
solutions to a four-dimensional low energy limit of string theory
containing non-minimally interacting electromagnetic and dilaton
fields. Some explicit homogeneous and inhomogeneous cosmological
solutions are constructed. For example, inhomogeneous exact
solutions presenting Gowdy - type EMD universe are obtained. The
asymptotic behaviour of the solutions is investigated. The
asymptotic form of the metric near the initial singularity  has a
spatially varying Kasner form. The character of the space-time
singularities is discussed. The late evolution of the solutions is
described by a background homogeneous and anisotropic universe
filled with weakly interacting gravitational, dilatonic and
electromagnetic waves.

\end{abstract}

\vskip 0.8cm


\vskip2pc]

\newcommand{\lfrac}[2]{{#1}/{#2}}
\newcommand{\sfrac}[2]{{\small \hbox{${\frac {#1} {#2}}$}}}

\section{Introduction}

In the last decade string cosmologies attracted large amount of
interest (see e.g. \cite{LWC} and references therein). In the
traditional approach the present (low energy) string cosmology is
in fact the classical cosmology where general relativity is
generalized by including additional (in most cases) massless
scalar fields. In general, one expects that the inclusion of these
extra matter degrees of freedom with the corresponding physical
interpretation, may somehow resolve the long standing problems in
cosmology. A recent interesting and promising development in this
direction is the so called {\em pre-big bang} scenario \cite{V1}.
In the framework of this scenario one assumes that the initial
state of the universe is characterized by small string coupling
and small curvature. This leads to an inflationary phase for
sufficiently homogeneous initial conditions. There are, however,
no natural reasons for the early universe not be inhomogeneous. At
present it is not clear how the large inhomogeneities may
influence the pre-big bang scenario. In a more general setting,
the question of whether our universe (which looks now homogeneous
and isotropic) may arise out of generic initial data still lacks a
complete answer in both general relativistic and string cosmology.

That is why the construction and study of exact inhomogeneous and
anisotropic  string-cosmological solutions are still of great
importance.

The inhomogeneous string cosmologies have been studied  by a
number of authors. In \cite{BK}, Barrow and Kunze have presented
inhomogeneous and anisotropic cosmological solutions of a low
energy string theory containing dilaton and axion fields when the
space-time metric possesses cylindrical symmetry . Their solutions
describe  ever-expanding universes with an initial curvature
singularity. The asymptotic form of the solutions near the initial
singularity has a spatially varying Kasner-like form. In
\cite{FLVM}, Fienstein, Lazkoz, and Vazquez - Mozo have presented
an algorithm for constructing exact solutions in string cosmology
for heterotic and type - IIB superstrings in four dimensions. They
have also presented and discussed some properties of an
inhomogeneous string cosmology with $S^{3}$ topology of the
spatial sections. Clancy et al. \cite{CFLT} have derived families
of anisotropic and inhomogeneous string cosmologies containing
nontrivial dilaton and axion fields by applying the global
symmetries of the string-effective action to a generalized
Eistein-Rosen metric. Lazkoz \cite{L} has presented an algorithm
for generating families of inhomogeneous space-times with a
massless scalar field. New solutions to Einstein - massless scalar
field equations having single isometry, have been generated in
\cite{L} by breaking the homogeneity of massless scalar field
$G_{2}$-models along one direction.

It should be noted, however, that the inhomogeneous cosmologies in
the framework of general relativity with certain physical fields
as a matter source have been investigated long before the time of
the string cosmology. Exact inhomogeneous vacuum Einstein
cosmologies with $S^{1}\times S^{1}\times S^{1}$, $S^{2}\times
S^{1}$ and $S^{3}$ topology of spatial sections have been found by
Gowdy \cite{Gowdy} and studied afterwards by Berger \cite{Berger}
and Misner \cite{Misner}. Exact stiff perfect fluid inhomogeneous
cosmologies have been studied by Wainwright, Ince and Marshman
\cite{WIM}. Later Charach \cite{C} and then Charach and Malin
\cite{CM} (see also \cite{CCM}) have found and studied exact
inhomogeneous cosmological solutions to the Einstein equations
with an electromagnetic and minimally coupled scalar field. These
solutions have been interpreted as an inhomogeneous universe
filled with gravitational, scalar and electromagnetic waves.

The purpose of this paper is to present sufficiently general
solution generating techniques  which permit to construct exact
inhomogeneous and anisotropic solutions to the equations of low
energy string theory containing non-minimally coupled dilaton and
electromagnetic fields - the so called Einstein-Maxwell-dilaton
(EMD) gravity. Examples of both homogeneous and inhomogeneous
exact solutions will be also presented and their asymptotics will
be investigated.

Besides the fact that  the EMD cosmologies are interesting in
their own, there are at least two more motivations for considering
the EMD cosmologies. First, a long standing question in cosmology
is the existence of a primordial magnetic field. The string theory
prediction of the non-minimal coupling of the dilaton to the
Maxwell  field  gives an opportunity to study the problem in a
more general dynamical context. Recently, in the homogeneous case
this question has been addressed in \cite{Giovannini}.

The second motivation is to investigate the character of initial
cosmological singularities in the presence of matter fields
arising from the low energy super-string theory. According to the
Belinskii-Khalatnikov-Lifhsitz (BKL) conjecture the dynamics of
nearby observers would decouple near the singularity for different
spatial points. A vacuum spatially homogeneous space-time
singularity (IX Bianchi type) is described by BKL as an infinite
sequence of Kasner epochs ("oscillatory or mixmaster
behaviour")\cite{BKL}, \cite{Misner1}. BKL further speculated that
a generic singularity should exhibit such a local oscillatory
behaviour.

Interesting special cases are the so-called asymptotic
velocity-term dominated (AVTD) singularities \cite{ELS}. Their
characteristic feature is the fact that the spatial derivative
terms in the field equations are negligible sufficiently close to
the singularities. Such singularities are described by only one
Kasner epoch, not by an infinite sequence of Kasner epochs i.e. an
oscillatory behaviour does not exist. We think that it is
important to study the behaviour of the singularities in the
presence of matter fields coming form the super-string theory. In
particular, it is interesting to investigate the influence of
non-minimal exponential coupling of the dilaton to the $U(1)$-
field on the character of the singularities. As it has been shown
in \cite {BKh},\cite{Berger1} the minimally coupled scalar field
can suppress the oscillatory behaviour.

Very recently the influence of the exponential
dilaton-electromagnetic coupling  on the character of initial
singularities has been studied by Narita, Torii and Maeda in
\cite{NTM}. These authors have considered $T^3$ Gowdy cosmologies
in Einstein-Maxwell-dilaton-axion system and have shown using the
Fushian algorithm that space-times in general have asymptotic
velocity-term dominated singularities. Their results mean that the
exponential coupling of the dilaton to the Maxwell field  does not
change the nature of the singularity. It should be also noted that
the exact solutions found in the present work lead to the same
conclusion.

Although, in the present paper we shall not consider the BKL
conjecture in details, we believe that the present work may serve
as a good ground for further studying of the above mentioned
questions in the framework of Einstein - Maxwell-dilaton-axion
gravity.

\section{Constructing solutions with one Killing vector}

We consider EMD-gravity described by the action

$$ {\cal A}= {1\over 16\pi} \int \left ({\star \,R} - 2d\varphi
\wedge {\star\, d\varphi}- 2e^{-2\varphi}F\wedge {\star\,
F}\right) $$

where $\star$ is Hodge dual with respect to the space-time metric
$g$, $R$ is the Ricci scalar curvature, $\varphi$ is the dilaton
field and $F=dA$ is the Maxwell two form.

First we will examine the case of space-time admitting one
space-like, hyper-surface orthogonal Killing vector $X$.
Physically this situation corresponds to a universe with
homogeneity  broken along two space directions. In the presence of
a Killing vector the Bianchi identity $dF=0$ and the Maxwell
equations $d\star \,e^{-2\varphi}F=0$ give rise to the following
local potentials $\Psi_{e}$ and $\Psi_{m}$:

$$ d\Psi_{e} = - i_{X}F   \,\,\,\, , \,\,\,\, d\Psi_{m} =
e^{-2\varphi} i_{X}\star \,F  .$$

Here $i_{X}$ is the interior product of an arbitrary form  with
respect to $X$.

The hyper-surface orthogonal Killing vector $X$ naturally
determines a three dimensional space-time submanifold $ \Sigma $
with metric \cite{Geroch}

$$ P= {\cal N} g - X\otimes X $$

where ${\cal N}= g(X,X)$ is the norm of the Killing vector.

We note that throughout the text we denote the Killing fields and
their naturally corresponding one-forms by the same symbols.

The projection on $\Sigma$ of the Einstein equations reads

\begin{eqnarray} \label{PEE}  {\hat {\cal R}} = 2d\varphi\otimes d\varphi +
{1\over 2 {\cal N}^2}d{\cal N}\otimes d{\cal N} + \nonumber \\
{2\over {\cal N}}\left(e^{-2\varphi}d\Psi_{e}\otimes d\Psi_{e} +
e^{2\varphi}d\Psi_{m}\otimes d\Psi_{m} \right) \end{eqnarray}

where $\hat {\cal R}$ is Ricci tensor with respect to the
projection metric $P$. The Maxwell equations take the form

\begin{equation} \label{PME}  d^{\dagger}\left(e^{-2\varphi}{\cal
N}^{-1}d\Psi_{e} \right)= 0  \,\,\,\,\,\,\,\, ,\,\,\,\,\,\,\,\,
d^{\dagger}\left(e^{2\varphi}{\cal N}^{-1}d\Psi_{m} \right)= 0  .
\end{equation}

Here $d^{\dagger}= \star\, d\, \star$\,  is the co-derivative
operator.

The projection of the Einstein equations along to the Killing
vector gives

\begin{eqnarray} \label{NKV}   {\cal N}\Box {\cal N}- g(d{\cal N}, d{\cal
N})= \nonumber
\\ -2{\cal N}e^{-2\varphi}\left(
e^{-2\varphi}g(d\Psi_{e},d\Psi_{e}) +
e^{2\varphi}g(d\Psi_{m},d\Psi_{m})\right) .\end{eqnarray}

Finally, the vanishing of the twist ${\cal T} = {1\over
2}\star\,dX$ of the Killing field leads to the constraint

\begin{equation} \label{CON}  d\Psi_{e}\wedge d\Psi_{m} = 0 .\end{equation}

The equations (\ref{PEE}), (\ref{PME}), (\ref{NKV}) and
(\ref{CON}) are equivalent to the EMD - gravity equations in the
presence of a spice-like, hyper-surface orthogonal Killing vector.

The constraint (\ref{CON}) implies that the one-forms $d\Psi_{m}$
and $d\Psi_{e}$ are proportional i.e. there exists a function
$\mathcal{F}$ such that $d\Psi_{m}={\mathcal{F}}\, d\Psi_{e}$.
Here we will consider the simplest case in which one of the
potentials $\Psi$ vanishes, say $\Psi_{e}= 0$.

In the case $\Psi_{e}=0$ the full system dimensionally reduced
equations can be obtained from the following action

\begin{eqnarray} \label{DRA} {\cal A_{R}}= {1\over 16\pi}\int \sqrt{-\det
P}\left({\cal R}
  - {P^{ij}\over 2{\cal N}^2}D_{i}{\cal N}D_{j}{\cal N}
    \right.  \nonumber \\
\left.    - P^{ij}D_{i}\varphi D_{j}\varphi - 2 e^{2\varphi}{\cal
N}^{-1}P^{ij}D_{i}\Psi_{m}D_{j}\Psi_{m} \right) .\end{eqnarray}

Now we introduce the following symmetric matrix $S \in
GL^{+}(2,R)$ :

$$ S= \left( \begin{array}{cc} {\cal N} +
2\Psi^{2}_{m}e^{2\varphi}& \sqrt{2}\Psi_{m}e^{2\varphi} \\& \\
\sqrt{2}\Psi_{m}e^{2\varphi} & e^{2\varphi}\\ \end{array} \right)
. $$

Making use of the matrix $S$, we may write the action (\ref{DRA})
in the form

\begin{equation} \label{MDRA} {\cal A_{R}}= {1\over 16\pi}\int\! \sqrt{-\det
P}\left({\cal R}
  - {1\over 2}P^{ij}\,Tr(D_{i}S D_{j}S^{-1}) \right) .
\end{equation}

The action (\ref{MDRA}) has $GL(2,R)$ as  a group of global
symmetry for fixed projection metric $P$. Explicitly the group
$GL(2,R)$ acts as follows :

$$ S\to ASA^{T} $$

where $A\in GL(2,R)$.

Note, that in the case when one of the electromagnetic potentials
vanishes the dimensionally reduced EMD equations can be viewed as
three-dimensional Einstein gravity coupled to a nonlinear $\sigma$
- model on the factor $GL(2,R)/O(2,R)$.

The $GL(2,R)$ symmetry may be employed for generating new exact
solutions from known ones. In particular, it will be very useful
to employ the $GL(2,R)$ symmetry to generate exact solutions with
nontrivial electromagnetic field from any given solution to the
vacuum field equations (i.e. pure Einstein equations) or the
dilaton-vacuum equations (Einstein equations plus a dilaton
field).

For example, starting with a dilaton-vacuum seed solution

$$ S_{vd}= \left( \begin{array}{cc} {\cal N}_{vd} & 0 \\& \\ 0 &
e^{2\varphi_{vd}}\\ \end{array} \right) ,$$

the nonlinear action of $GL(2,R)$ gives

$$ {\cal N} = (\det A)^2 {{\cal N}_{vd}\, e^{2\varphi_{vd}} \over
c^2 {\cal N}_{vd} + d^2 e^{2\varphi_{vd}}} ,$$

$$ e^{2\varphi} =  c^2 {\cal N}_{vd} + d^2 e^{2\varphi_{vd}} ,$$

$$ \Psi_{m} = {1\over \sqrt{2}}\,\, { ac {\cal N}_{vd} + bd
e^{2\varphi_{vd}}  \over c^2 {\cal N}_{vd} + d^2 e^{2\varphi_{vd}}
} $$

where

$$ A= \left( \begin{array}{cc} a & b \\ c & d\\ \end{array}
\right). $$

It is worth noting, however, that some of the $GL(2,R)$-
transformations are pure electromagnetic gauge or rescaling of the
solutions. Only the $O(2,R)$-transformations lead to essentially
new solutions.

Exact solutions to the EMD gravity equations can be constructed
from vacuum solutions by the  method described in
\cite{Yazadjiev}, too. In order to obtain solutions by this method
one assumes that the matrix $S$ depends on space-time coordinates
through a harmonic potential $\Omega$ ($D^{i}D_{i}\,\Omega =0$).
Requiring then the constraint $$-{1\over 4}Tr\left({dS\over
d\Omega}{dS^{-1}\over d\Omega}\right)=1$$ to be satisfied, the EMD
equations are reduced to the vacuum Einstein equations

\begin{eqnarray*} {\hat {\cal R}}_{ij}= 2D_{i}\Omega\, D_{j}\Omega
\; , \nonumber \\ D^{i}D_{i}\Omega = 0 \end{eqnarray*}

and a separated matrix equation

$$ {d\over d\Omega}\left(S^{-1}{dS\over d\Omega}\right)= 0 \, . $$

Therefore, for every solution to the vacuum Einstein equations,
the solutions of the matrix equation give classes of exact EMD
gravity solutions (see for details \cite{Yazadjiev}).

It should be noted that the seed solutions used for the
construction of exact EMD solutions via the above described
methods, must admit at least one Killing vector. The case with two
commuting hyper-surface orthogonal Killing vectors will be
considered in the following section.

\section{Constructing solutions with two Killing vectors}

\subsection{General equations}

In this section we consider the EMD gravity equations when the
space-time admits two commuting  space-like, hyper-surface
orthogonal Killing vectors $X$ and $Y$. In this case the metric
can be written in the form

\begin{equation} \label{RM}  ds^2 = e^{2h - 2p}(dz^2 - d\xi^2) + e^{2p}dx^2 +
\varrho \,\, e^{-2p}dy^2 \end{equation}

where $h$, $p$ and $\varrho$ are unknown functions  of $z$ and
$\xi$ only. Therefore the Killing vectors are given by $X =
{\partial \over \partial x}$ $Y = {\partial \over \partial y} $.

The physical properties of the metric (\ref{RM}) depend on the
gradient of the norm $\varrho$ of the two-form $X \wedge Y$ . The
case corresponding to $\partial_{\mu}\varrho$ being globally
space-like or null describes cylindrical and  plane gravitational
waves. Metrics where the sign of $\partial_{\mu}\varrho \,
\partial^{\mu}\varrho$ can change, describe
colliding plane waves or cosmological models with space-like and
time-like singularities. Metrics with globally time-like
$\partial_{\mu}\varrho$ describe cosmological models with
space-like singularities.

In the present paper we shall consider only the globally time-like
case $\varrho = \xi^2$. It should be noted that the choice
$\varrho = \xi^2$ is dynamically consistent with the EMD gravity
equations because the Ricci tensor ${\hat R}$ satisfies

$$ g(Y,Y){\hat R}(X,X) + g(X,X){\hat R}(Y,Y) = 0 . $$

In the presence of a second Killing vector the dimensional
reduction can be further continued. Without going into details we
directly present the dimensionally reduced equations

\begin{eqnarray} \label{EP}
\partial^2_{\xi}p + {1\over \xi }\partial_{\xi}p - \partial^2_{z}p
= e^{-2\varphi - 2p}\left((\partial_{z}\Psi_{e})^2 -
(\partial_{\xi}\Psi_{e})^2 \right) \nonumber \\ +
\,\,\,\,e^{2\varphi - 2p}\left((\partial_{z}\Psi_{m})^2 -
(\partial_{\xi}\Psi_{m})^2 \right) , \end{eqnarray}

\begin{eqnarray} \label{EF}
\partial^2_{\xi}\varphi + {1\over \xi }\partial_{\xi}\varphi -
\partial^2_{z}\varphi
= e^{-2\varphi - 2p}\left((\partial_{z}\Psi_{e})^2 -
(\partial_{\xi}\Psi_{e})^2 \right) \nonumber  \\ -
\,\,\,\,e^{2\varphi - 2p}\left((\partial_{z}\Psi_{m})^2 -
(\partial_{\xi}\Psi_{m})^2 \right) ,\end{eqnarray}

\begin{equation} \label{CON1}
\partial_{\xi}\Psi_{e} \partial_{z}\Psi_{m} =
\partial_{z}\Psi_{e} \partial_{\xi}\Psi_{m} ,
\end{equation}

\begin{equation} \label{EPSIE}
\partial_{\xi}\left(\xi e^{-2\varphi - 2p}\partial_{\xi}\Psi_{e}
\right) -
\partial_{z}\left(\xi e^{-2\varphi - 2p}\partial_{z}\Psi_{e}
\right) = 0  , \end{equation}

\begin{equation} \label{EPSIM}
\partial_{\xi}\left(\xi e^{2\varphi - 2p}\partial_{\xi}\Psi_{m}
\right) -
\partial_{z}\left(\xi e^{2\varphi - 2p}\partial_{z}\Psi_{m}
\right) = 0 ,\end{equation}

\begin{eqnarray} \label{EHKSI} {1\over \xi}\partial_{\xi}h  =
(\partial_{\xi}\varphi)^{2} + (\partial_{z}\varphi)^{2} +
(\partial_{\xi}p)^{2} + (\partial_{z}p)^{2} + \nonumber \\
e^{-2\varphi - 2p}\left((\partial_{\xi}\Psi_{e})^2  +
(\partial_{\xi}\Psi_{e})^2 \right)  + \nonumber \\  e^{-2\varphi -
2p}\left((\partial_{\xi}\Psi_{m})^2 +  (\partial_{\xi}\Psi_{m})^2
\right)  , \end{eqnarray}

\begin{eqnarray} \label{EHZ} {1\over \xi} \partial_{z}h =
2\partial_{\xi}\varphi\partial_{z}\varphi \,\, +
2\partial_{\xi}p\partial_{z}p  + \nonumber \\ 2e^{-2\varphi -
2p}\partial_{\xi}\Psi_{e}\partial_{z}\Psi_{e} + \,\, 2e^{2\varphi
- 2p}\partial_{\xi}\Psi_{m}\partial_{z}\Psi_{m} \;.\end{eqnarray}

Here we should make some comments. We have written the system of
partial differential equations in terms of the electromagnetic
potentials $\Psi_{e}$ and $\Psi_{m}$. These potentials are derived
with respect to the Killing vector $X$. In the same way we may use
the electromagnetic potentials with respect to the Killing vector
$Y$ denoted respectively as $\Psi^{Y}_{e}$ and $\Psi^{Y}_{m}$.
Moreover, we may also use  a mixed pair of potentials, say
$\Psi_{e}$ and $\Psi^{Y}_{e}$. In general, the transition between
different pairs of potentials may be performed by the following
formulas $$
\partial_{\xi}\Psi^{Y}_{m} = - \,\xi \, e^{-2\varphi -
2p}\,\,\partial_{z}\Psi_{e} , $$

$$ \partial_{z}\Psi^{Y}_{m}= -\,\xi \, e^{-2\varphi -
2p}\,\,\partial_{\xi}\Psi_{e} , $$

$$
\partial_{z}\Psi^{Y}_{e}= - \,\xi \, e^{2\varphi -
2p}\,\,\partial_{\xi}\Psi_{m}  ,$$ $$
\partial_{\xi}\Psi^{Y}_{e} = - \, \xi \, e^{2\varphi -
2p}\,\,\partial_{z}\Psi_{m}  .$$

For the reader's convenience we have presented in the Appendix the
system (\ref{EP}) - (\ref{EHZ}) written in terms of the mixed pair
$\Psi_{e}=\omega$ and $\Psi^{Y}_{e}=\chi$ .

Let us go back again to the system of nonlinear partial
differential equations (\ref{EP}) - (\ref{EHZ}). We have a
complicated system of coupled nonlinear partial differential
equations and it seems that finding all its solutions is a
hopeless task. Nevertheless, as we will see a large enough class
of solutions can be found.

\subsection{Solution generating method}

Here we consider the case in which one of the electromagnetic
potentials vanishes. For definiteness we take $\Psi_{e}=0$. Let us
introduce the new potentials $u=p - \varphi$, $\phi = p +
\varphi$, $\Psi^{S}_{m}= \sqrt{2}\Psi_{m}$ and $h^{S}=2h$. Then
the system (\ref{EP}) - (\ref{EHZ}) may be rewritten in terms of
the new variables as follows

$$  \partial^2_{\xi}\phi + {1\over \xi }\partial_{\xi}\phi -
\partial^2_{z}\phi
= 0 ,$$ $$
 \partial^2_{\xi}u + {1\over \xi }\partial_{\xi}u -
\partial^2_{z}u = e^{-2u}\left((\partial_{z}\Psi^{S}_{m})^2  -
 (\partial_{\xi}\Psi^{S}_{m})^2 \right) ,
$$

\begin{equation} \label{SMCSF}
\partial_{\xi}\left(\xi e^{-2u} \partial_{\xi}\Psi^{S}_{m}\right)
- \partial_{z}\left(\xi e^{-2u}
\partial_{z}\Psi^{S}_{m}\right) = 0 ,
\end{equation}

\begin{eqnarray*}
{1\over \xi}\partial_{\xi}h^{S}= (\partial_{\xi}\phi)^2  +
(\partial_{z}\phi)^2 + (\partial_{\xi}p^{S})^2 +
(\partial_{z}p^{S})^2 +  \\ e^{-2u}
\left((\partial_{\xi}\Psi^{S}_{m})^2 +
(\partial_{z}\Psi^{S}_{m})^2 \right)  ,\end{eqnarray*}

$$ {1\over \xi}\partial_{z}h^{S}= 2\partial_{\xi}\phi
\partial_{z}\phi +  2\partial_{\xi}u
\partial_{z}u  + 2e^{-2u}\partial_{\xi}\Psi^{S}_{m}
\partial_{z}\Psi^{S}_{m} .
$$

It is not difficult to recognize that the system (\ref{SMCSF})
coincides with the corresponding one for the Einstein-Maxwell
gravity with a minimally coupled scalar field $\phi$ for the
metric

\begin{equation} \label{RM1}  ds^2 = e^{2h^{S}- 2u}(dz^2 - d\xi^2) +
e^{2u}dx^2 + e^{-2u}\xi^2 dy^2  . \end{equation}

Here $\Psi^{S}_{m}$ is the corresponding non-vanishing
electromagnetic potential in the case of a minimally coupled
scalar field.

In this way we have proven the following proposition

\begin{prop}\label{Proposition}

 Let $u$, $\phi$, $h^{S}$ and $\Psi^{S}_{m}$ be a solution to the
Einstein-Maxwell equations with a minimally coupled scalar field
for the metric (\ref{RM1}). Then $p= {1\over 2}(u + \phi) $,
$\varphi = {1\over 2}(\phi - u)$, $h= {1\over 2}h^{S}$ and
$\Psi_{m}= {1\over \sqrt{2}}\Psi^{S}_{m}$ form a solution to the
equations of EMD gravity for the metric (\ref{RM}).\end{prop}

This result allows us to generate exact solutions to the EMD
gravity equations for the metric (\ref{RM}) directly from known
solutions of the Einstein - Maxwell equations with a minimally
coupled scalar field.

Before going further we shall make some comments. In the case when
the space-time admits two space-like Killing symmetries and one of
the electromagnetic potentials is taken to vanish, it may be shown
that the equations governing the transverse part of the
gravitational field, dilaton and non-vanishing electromagnetic
potential are written in the compact matrix form

\begin{equation} \label{CHIRAL}
\partial_{\xi}\left(\xi S^{-1}\partial_{\xi}S\right)
  - \partial_{z}\left(\xi S^{-1}\partial_{z}S \right)=
  0 . \end{equation}

The equation (\ref{CHIRAL}) is just the chiral equation. There are
powerful solitonic techniques for solving this equation (see
\cite{BZ}). Although, the study of cosmological soliton solutions
in EMD gravity seems to be very interesting, we will not consider
them in the preset work.

\section{Examples of exact solutions}

\subsection{Homogeneous solutions}

Homogeneous solutions are obtained when we have dependence only on
the "time - coordinate" $\xi$. In this case the differential
constraint reduces to
$\partial_{\xi}\Psi_{e}\partial_{\xi}\Psi_{m}=0$. Thus we may
choose one of the potentials to vanish. Therefore we may apply the
proposition from the previous section to generate homogeneous
solutions starting from the corresponding homogeneous solutions
for the case of a minimally coupled scalar field. In our notations
the homogeneous solutions for a minimally coupled scalar field are
\cite{CCM}

\begin{equation} \label{HSS1} \phi = \phi_{0} + \beta_{0}\ln(\xi)
,
\end{equation}

\begin{equation} \label{HSS2} u = \ln \left( e^{\phi_{1}\over
2}\xi^{1+{\alpha_{0}\over 2}} + e^{-\phi_{1}\over 2}\xi^{1 -
{\alpha_{0}\over 2}} \right), \end{equation}

\begin{equation} \label{HSS3} \Psi^{SY}_{e}= {1\over 2} \tanh ({\phi_{1}\over
2} + {\alpha_{0}\over 2}\ln(\xi)) , \end{equation}

where $\phi_{0}$, $\phi_{1}$, $\alpha_{0}$ and $\beta_{0}$ are
arbitrary constants.

The above seed solution (\ref{HSS1}) - (\ref{HSS3}) gives the
following homogeneous solution to the EMD equations. The potential
$\Psi^{Y}_{e}$ is

$$ \Psi^{Y}_{e} = {1\over 2\sqrt{2}} \tanh ({\phi_{1}\over 2} +
{\alpha_{0}\over 2}\ln(\xi)) .$$

The metric has the form

$$ ds^2= A^2(\xi)(dz^2 - d\xi^2) + B^{2}(\xi)dx^2 + C^{2}(\xi)dy^2
$$

where

$$ A^{2}(\xi)= e^{\gamma_{0} - \phi_{0}}\xi^{ {\alpha_{0}^2 \over
4} + \beta_{0}^2 - \beta_{0}} \left(e^{\phi_{1}\over
2}\xi^{{\alpha_{0}\over 2}} + e^{-\phi_{1}\over 2}\xi^{ -
{\alpha_{0}\over 2}} \right) , $$

$$ B^{2}(\xi) = e^{\phi_{0}}\xi^{1 + \beta_{0}}
\left(e^{\phi_{1}\over 2}\xi^{{\alpha_{0}\over 2}} +
e^{-\phi_{1}\over 2}\xi^{ - {\alpha_{0}\over 2}} \right) ,$$

$$ C^{2}(\xi) = e^{-\phi_{0}}\xi^{1 - \beta_{0}}
\left(e^{\phi_{1}\over 2}\xi^{{\alpha_{0}\over 2}} +
e^{-\phi_{1}\over 2}\xi^{ - {\alpha_{0}\over 2}} \right)^{-1} .$$

Here $\gamma_{0}$ is a constant.

The asymptotic behaviour of the expansion factors and the dilaton
field  in the limit $\xi \to 0$ is as follows

$$ A^{2}(\xi) \sim  \xi^{ {\alpha_{0}^2 \over 4} + \beta_{0}^2 -
\beta_{0} - {\mid \alpha_{0}\mid \over 2}} , $$

$$ B^{2}(\xi) \sim \xi^{ 1 + \beta_{0} - {\mid \alpha_{0}\mid
\over 2}} ,$$

$$ C^{2}(\xi) \sim \xi^{ 1 - \beta_{0} + {\mid \alpha_{0}\mid
\over 2}} , $$

$$ \varphi \sim -{1\over 2}(1 - \beta_{0} - {1\over 2}\mid
\alpha_{0}\mid ) \ln(\xi) .$$

The asymptotic form of the expansion factors and the dilaton field
 in the limit $\xi \gg 1$ is

$$ A^{2}(\xi) \sim  \xi^{ {\alpha_{0}^2 \over 4} + \beta_{0}^2 -
\beta_{0} + {\mid \alpha_{0}\mid \over 2}} ,$$

$$ B^{2}(\xi) \sim \xi^{ 1 + \beta_{0} + {\mid \alpha_{0}\mid
\over 2}} ,$$

$$ C^{2}(\xi) \sim \xi^{ 1 - \beta_{0} - {\mid \alpha_{0}\mid
\over 2}} ,$$

$$ \varphi \sim -{1\over 2}(1 - \beta_{0} + {1\over 2}\mid
\alpha_{0}\mid )\ln(\xi) .$$

Introducing the scynhroneous time $d\tau = A(\xi)d\xi$, the line
element in the limits $\xi\to 0$ and $\xi \gg 1$ can be written in
the Kasner form

$$ ds^2 \sim -d\tau^2 + \tau^{2p_{1}}dx^2 + \tau^{2p_{2}}dy^2 +
\tau^{2p_{3}}dz^2 $$

where the Kasner exponents are defined by

$$ p_{1}= {1 + \beta_{0} \mp {1\over 2} \mid\alpha_{0}\mid \over
{1\over 4}(1 \mp   \mid\alpha_{0}\mid )^2  + ({1\over 2} -
\beta_{0})^2 + {3\over 2}} \; ,$$

$$p_{2}= {1 - \beta_{0} \pm {1\over 2} \mid\alpha_{0}\mid \over
{1\over 4}(1 \mp \mid\alpha_{0}\mid )^2  + ({1\over 2} -
\beta_{0})^2 + {3\over 2}} \; ,$$

$$ p_{3}= {{1\over 4}(1 \mp   \mid\alpha_{0}\mid )^2  + ({1\over
2} - \beta_{0})^2 - {1\over 2} \over  {1\over 4}(1 \mp
\mid\alpha_{0}\mid )^2  + ({1\over 2} - \beta_{0})^2 + {3\over
2}}\; , $$

as the sing "$-$" refers to the limit $\xi \to 0$ while the sign
"$+$" is for the limit $\xi \gg 1$.

The dilaton filed is given by

$$ \varphi \sim \sigma \ln(\tau) $$

where

$$ \sigma = -{1 - \beta_{0} \mp {1\over 2}\alpha_{0} \over {1\over
4}(1 \mp \mid\alpha_{0}\mid )^2  + ({1\over 2} - \beta_{0})^2 +
{3\over 2}} \; .$$

The parameters $p_{1}$, $p_{2}$ and  $p_{3}$ satisfy the Belinskii
-Khalatnikov  relations

\begin{eqnarray*} &&p_{1} + p_{2} + p_{3} = 1  ,\\ &&p_{1}^2 + p_{2}^2 +
p_{3}^2 = 1 - 2\sigma^2  \, . \end{eqnarray*}

\subsection{Inhomogeneous solutions}

Here the proposition (\ref{Proposition}) will be applied again,
this time to generate  inhomogeneous exact solutions to the EMD
equations. As a seed family of solutions we take the Charach
family of solutions describing inhomogeneous cosmologies with
minimally coupled scalar and electromagnetic fields and with
$S^{1}\times S^{1}\times S^{1}$ topology of the spatial sections.
The Charach family of inhomogeneous solutions and their
asymptotics in the two limiting cases are presented in the
Appendix.

According to the proposition (\ref{Proposition}) the inhomogeneous
EMD metric is given by

$$ ds^2 = e^{2(h - p)}(dz^2 - dt^2) + e^{2p}dx^2 + \xi^2
e^{-2p}dy^2 $$

where

$$ 2p = \ln(2\xi\cosh({1\over 2}{\tilde \phi})) + \phi ,$$

\begin{eqnarray*} h= {1\over 2}\ln(\xi) + \ln(2\cosh({1\over
2}{\tilde\phi})) + \\ {1\over 4}F({\tilde
\phi_{0}},\alpha_{0},A_{n}, B_{n};\xi,z) + F(
\phi_{0},\beta_{0},C_{n}, D_{n};\xi,z) .\end{eqnarray*}

The dilaton field is

$$ \varphi = {1\over 2} \phi - {1\over
2}\ln\left(2\xi\cosh({1\over 2}{\tilde \phi}) \right) .$$

The inhomogeneous cosmological solutions introduce characteristic
length scales. In fact each normal mode has its own characteristic
scale. The horizon distance in the $"z"$ direction is given by
$ds^2\mid_{x,y}=0$ and hence

$$ \delta z= \int_{0}^{\xi}d\xi =\xi .$$

In this way $n\xi$ can be viewed as the ratio of the horizon
distance in the $"z"$ direction  to the coordinate wavelength
$\lambda_{n}$ i.e. $n\xi= {\delta z \over \lambda_{n}}$.

There are two limiting cases to be considered. The first case is
when the wavelength is much larger than the horizon scale ($n\xi
\ll 1 $). The second case is when the wavelength is much less than
the horizon scale ($n\xi \gg 1$).

In the first case ($n\xi \ll 1$) the asymptotic form of the EMD
metric is

$$ ds^2 = A^{2}(\xi,z)(dz^2 - d\xi^2) + B^{2}(\xi,z)dx^2 +
C^{2}(\xi,z)dy^2 $$

where

\begin{eqnarray*} && A^{2}(\xi,z) =  \\ && e^{\gamma(z)}\xi^{{1\over
4}\alpha^{2}(z)+ \beta^{2}(z) + \beta(z)}\!\!\left(e^{{1\over 2}
{\tilde \phi}_{*}(z)}\xi^{{1\over 2}\alpha(z)} + e^{-{1\over 2}
{\tilde \phi}_{*}(z)}\xi^{-{1\over 2}\alpha(z)} \right) ,
\end{eqnarray*}
\begin{eqnarray*}
 && B^{2}(\xi,z) = \nonumber \\ && e^{{\tilde \phi}_{*}(z)}\xi^{1 + \beta(z)}
 \left(e^{{1\over 2}
{\tilde \phi}_{*}(z)}\xi^{{1\over 2}\alpha(z)} + e^{-{1\over 2}
{\tilde \phi}_{*}(z)}\xi^{-{1\over 2}\alpha(z)} \right) ,
\end{eqnarray*}
\begin{eqnarray*} && C^{2}(\xi,z) =  \\ && e^{-{\tilde
\phi}_{*}(z)}\xi^{1 - \beta(z)}
 \left(e^{{1\over 2}
{\tilde \phi}_{*}(z)}\xi^{{1\over 2}\alpha(z)} + e^{-{1\over 2}
{\tilde \phi}_{*}(z)}\xi^{-{1\over 2}\alpha(z)} \right)^{-1} .
\end{eqnarray*}

In the limit $n\xi \ll 1$ the asymptotic form of the dilaton is

\begin{eqnarray*} \varphi \sim {1\over 2}\left(\phi_{*}(z) +
sign(\alpha(z)){\tilde \phi_{*}(z) } \right)- \\{1\over
2}\left(1-\beta(z)-{1\over 2}\mid \alpha(z)\mid \right)\ln(\xi) .
\end{eqnarray*}

In order to discuss the cosmological solutions near the
singularity we have to consider homogeneous limit for $\xi$
approaching zero. Introducing the proper time, in the homogeneous
limit, by $$\tau = \int A(\xi)d\xi ,$$ we find that the metric has
the Kasner form

$$ g_{\mu\nu} \sim (-1,\tau^{2p_{1}},\tau^{2p_{2}},\tau^{2p_{3}})
.$$

Here the Kasner indexes are spatially varying

$$ p_{1}(z)= {1 + \beta(z) \mp {1\over 2} \mid\alpha(z)\mid \over
{1\over 4}(1 \mp   \mid\alpha(z)\mid )^2  + ({1\over 2} -
\beta(z))^2 + {3\over 2}},$$

$$ p_{2}(z)= {1 - \beta(z) \pm {1\over 2} \mid\alpha(z)\mid \over
{1\over 4}(1 \mp \mid\alpha(z)\mid )^2  + ({1\over 2} -
\beta(z))^2 + {3\over 2}} ,$$

$$ p_{3}(z)= {{1\over 4}(1 \mp   \mid\alpha(z)\mid )^2  + ({1\over
2} - \beta(z))^2 - {1\over 2} \over  {1\over 4}(1 \mp
\mid\alpha(z)\mid )^2  + ({1\over 2} - \beta(z))^2 + {3\over 2}}
$$

and satisfy the Belinski- Khalatnikov relations

$$ p_{1}(z) + p_{2}(z) + p_{3}(z)= 1 ,$$ $$ p^2_{1}(z) +
p^2_{2}(z) + p^2_{3}(z) = 1 - 2\sigma^2(z) , $$

where $$ \sigma = -{1 - \beta(z) \mp {1\over 2}\alpha(z) \over
{1\over 4}(1 \mp \mid\alpha(z)\mid )^2  + ({1\over 2} -
\beta(z))^2 + {3\over 2}} .$$

In the second case when $n\xi \gg 1$ the asymptotic form of the
EMD metric is as follows.

The first sub-case  is for $\alpha_{0}=0$. Then the EMD metric is
given by

$$ g_{\mu\nu} = \eta_{\mu\nu} + h_{\mu\nu} $$

where

\begin{eqnarray*} \eta = diag\left(- 2\xi^{\beta^2_{0} - \beta_{0}}e^{{1\over
2 }K\xi}, 2\xi^{1 + \beta_{0}}, \right. \\ \left. {1\over 2}\xi^{1
- \beta_{0}}, 2\xi^{\beta^2_{0} - \beta_{0}}e^{{1\over 2 }K\xi}
\right) \end{eqnarray*}

and

$$ h = diag\left(0, {2\over \sqrt{\xi}}\xi^{1 +
\beta_{0}}H(\xi,z), -{2\over \sqrt{\xi}}\xi^{1 -
\beta_{0}}H(\xi,z), 0\right) .$$

When $\alpha_{0}\ne 0$ we have

\begin{eqnarray*} \eta = diag \left(- \xi^{{1\over 4}(1 + \alpha_{0})^2 +
(\beta_{0} - {1\over 2})^2  - {1\over 2}}e^{{1\over 2}K\xi},
\xi^{1 + {1\over 2}\mid\alpha_{0}\mid + \beta_{0}},  \right. \\
\left. \xi^{1 - {1\over 2}\mid\alpha_{0}\mid - \beta_{0}},
\xi^{{1\over 4}(1 + \alpha_{0})^2 + (\beta_{0} - {1\over 2})^2  -
{1\over 2}}e^{{1\over 2}K\xi}\right) ,\end{eqnarray*}
\begin{eqnarray*}
 h&= diag\!\!\left(0, {\xi^{1 + {1\over 2}\mid\alpha_{0}\mid +
\beta_{0}} \over \sqrt{\xi}}(H(\xi,z)+ {1\over 2}
sign(\alpha_{0}){\tilde H}(\xi,z) ), \right. \\ \!\!\!\!\!\!\! &
\left. -{\xi^{1 - {1\over 2}\mid\alpha_{0}\mid - \beta_{0}} \over
\sqrt{\xi}}(H(\xi,z)+ {1\over 2} sign(\alpha_{0}){\tilde H}(\xi,z)
), 0 \right) .\end{eqnarray*}

The asymptotic form of the dilaton in the limit $n\xi \gg 1$, for
both $\alpha_{0}=0$ and $\alpha_{0}\ne 0$, is given by

\begin{eqnarray*} \varphi \sim  -{1\over 2}\left(1 - \beta_{0} + {1\over
2}\mid\alpha_{0}\mid \right)\ln(\xi) + \\ {1\over
\sqrt{\xi}}\left(H(\xi,z) + {1\over 2}sign(\alpha_{0}){\tilde
H}(\xi,z)\right) .\end{eqnarray*}

The explicit form of the solutions and their asymptotics allow us
to make some conclusions. The obtained solutions fall in the
category of asymptotic velocity-term dominated space-times. When
the singularity is approached the spatial derivatives become
negligible with comparison to the time derivatives. Sufficiently
close to the singularity the evolution at different spatial points
is decoupled and the metric is locally Kasner with spatially
dependent Kasner indices satisfying the Belinskii-Khalatnikov
relations.

The late evolution of the exact cosmological solutions is
described by a homogeneous, anisotropic  universe with
gravitational, scalar (dilatonic) and electromagnetic waves. The
non-minimal dilaton-electromagnetic exponential coupling
influences mainly the homogeneous background universe rather the
scalar and electromagnetic waves on that background. The former
may be considered as minimally coupled scalar and electromagnetic
waves up to higher orders of ${1\over \xi}$.

\section{Conclusion}

In this paper we have shown that it is possible to find exact
inhomogeneous and anisotropic cosmological solutions of low energy
string theory containing non-minimally interacting dilaton and
Maxwell fields - Einstein-Maxwell -dilaton gravity. First we have
considered space-times admitting one hyper-surface orthogonal
Killing vector. It has been shown that in the case of one
vanishing electromagnetic potential the dimensionally reduced
equations possess a group of global symmetry $GL(2,R)$. We have
described an algorithm for generating exact cosmological EMD
solutions starting from vacuum and dilaton-vacuum backgrounds by
employing the nonlinear action of the global symmetry group.  This
algorithm is especially useful for generating exact cosmological
EMD solutions with only one Killing vector, starting  with
$G_{1}$-dilaton-vacuum background. Another method which permits to
construct exact EMD solutions starting from solutions of the pure
Einstein equations has been briefly discussed, too.

In the case when the space-time admits two commuting,
hyper-surface orthogonal Killing vectors we have given a method
which allows exact inhomogeneous cosmological solutions to the EMD
equations to be generated from the corresponding solutions of the
Einstein-Maxwell equations with a minimally coupled scalar field.
Using this method,  as a particular case, we have obtained exact
cosmological homogeneous and inhomogeneous solutions to the EMD
equations. The initial evolution described by these solutions is
of a spatially varying Kasner form. The intermediate stage of
evolution occurs when the characteristic scales of the
inhomogeneities approach the scale of the particle horizon. This
stage is characterized by strongly interacting non-linear
gravitational, dilatonic and electromagnetic waves. The late
evolution of the cosmological solutions is described by a
background homogeneous and anisotropic universe filled with weakly
interacting gravitational, dilatonic and electromagnetic waves.

The cosmological solutions found in the present work fall in the
category of AVTD space-times. Near the singularity the dynamics at
different spatial points decouples and the metric has a spatially
varying Kasner form. Therefore, the non-minimal coupling of the
dilaton to the Maxwell field does not change  the nature of the
singularity. The solutions are sufficiently generic and therefore
we should expect that the $T^{3}$ space-times in EMD gravity have
AVTD singularities in general.

Finally we believe that the present paper will be a sufficiently
good background for studying similar problems in the more-general
case of Enstein-Maxwell-dilaton-axion gravity.

\bigskip


\section*{Acknowledgments} The author is deeply grateful to E.
Alexandrova for the support and encouragement. Without her help
the writing of this paper most probably would be impossible.

The author is also grateful to P. Fiziev for the support and
stimulating discussions.

The author would particularly like to thank V. Rizov for having
read critically manuscript and for useful suggestions.


\bigskip

\begin{appendix}

\section{Reduced EMD system in terms of the potentials $\Psi_{e}=\omega$
and $\Psi^{Y}_{e}=\chi $}

\begin{eqnarray*}
\partial^2_{\xi}p + {1\over \xi }\partial_{\xi}p - \partial^2_{z}p
= e^{-2\varphi - 2p}\left((\partial_{z}\omega)^2 -
(\partial_{\xi}\omega)^2 \right) \nonumber \\ + \,\,\,\,{1\over
\xi^2} \,e^{-2\varphi + 2p}\left((\partial_{\xi}\chi)^2 -
(\partial_{z}\chi)^2 \right) \end{eqnarray*}
\begin{eqnarray*}
\partial^2_{\xi}\varphi + {1\over \xi }\partial_{\xi}\varphi -
\partial^2_{z}\varphi
= e^{-2\varphi - 2p}\left((\partial_{z}\omega)^2 -
(\partial_{\xi}\omega)^2 \right) \nonumber  \\ - \,\,\,\,{1\over
\xi^2} \,e^{-2\varphi + 2p}\left((\partial_{\xi}\chi)^2 -
(\partial_{z}\chi)^2 \right) \end{eqnarray*}

$$
\partial_{\xi}\omega \,\partial_{\xi}\chi =
\partial_{z}\omega \,\partial_{z}\chi
$$

\begin{eqnarray*}
\partial_{\xi}^2\omega + {1\over \xi}\partial_{\xi}\omega - \partial_{z}^2 \omega =
\nonumber \\ 2(\partial_{\xi}\varphi +
\partial_{\xi}p)\partial_{\xi}\omega - 2(\partial_{z}\varphi +
\partial_{z}p)\partial_{z}\omega \end{eqnarray*}
\begin{eqnarray*}
\partial_{\xi}^2\chi - {1\over \xi}\partial_{\xi}\chi - \partial_{z}^2 \chi =
\nonumber \\ 2(\partial_{\xi}\varphi -
\partial_{\xi}p)\partial_{\xi}\chi - 2(\partial_{z}\varphi -
\partial_{z}p)\partial_{z}\chi \end{eqnarray*}
\begin{eqnarray*}  {1\over \xi}\partial_{\xi}h  =
(\partial_{\xi}\varphi)^{2} + (\partial_{z}\varphi)^{2} +
(\partial_{\xi}p)^{2} + (\partial_{z}p)^{2} + \nonumber \\
e^{-2\varphi - 2p}\left((\partial_{\xi}\omega)^2  +
(\partial_{\xi}\omega)^2 \right)  + \nonumber \\ {1\over \xi^2}
e^{-2\varphi + 2p}\left((\partial_{\xi}\chi)^2 +
(\partial_{\xi}\chi)^2 \right)   \end{eqnarray*}
\begin{eqnarray*}
{1\over \xi} \partial_{z}h =
2\partial_{\xi}\varphi\partial_{z}\varphi \,\, +
2\partial_{\xi}p\partial_{z}p  + \nonumber \\ 2e^{-2\varphi -
2p}\partial_{\xi}\omega \partial_{z}\omega + \,\, 2e^{-2\varphi +
2p}\partial_{\xi}\chi \partial_{z}\chi \end{eqnarray*}

\section{Charach family of exact solutions}

For the reader's convenience we present here the Charach family of
solutions written in our notations.

The minimally coupled scalar filed is given by

\begin{eqnarray*} \phi= \phi_{0} +  \beta_{0}\ln(\xi) +  \\
\sum_{n=1}^{\infty}\left( C_{n}J_{0}(n\xi)  +
D_{n}N_{0}(n\xi)\right)\cos(n(z-z_{n})) .\end{eqnarray*}

Here $\phi_{0}$, $\beta_{0}$, $C_{n}$, $D_{n}$ are arbitrary
constants and $J_{0}(.)$ and $N_{0}(.)$ are respectively the
Bessel and Neumann functions.

The metric function $u$ and the non-vanishing electromagnetic
potential $\Psi^{SY}_{e}$ are correspondingly

$$ u = \ln\left(2\xi \cosh({1\over 2}{\tilde \phi}) \right), $$

$$ \Psi^{SY}_{e}= const + {1\over 2}\tanh({1\over 2}{\tilde \phi})
 $$

where ${\tilde \phi}$ is an auxiliary scalar field which is of the
form

\begin{eqnarray*} {\tilde \phi}= {\tilde \phi}_{0} + \alpha_{0}\ln(\xi)  +
\\ \sum_{n=1}^{\infty}\left( A_{n}J_{0}(n\xi)  +
B_{n}N_{0}(n\xi)\right)\cos(n(z-z_{n})) .\end{eqnarray*}

The longitudinal  part of the gravitational field is

\begin{eqnarray*} h^{S}= \ln(\xi) + 2\ln(2\cosh({1\over 2}{\tilde \phi})) +
\\ {1\over 2}F({\tilde \phi}_{0},\alpha_{0},A_{n},B_{n};\xi,z) +
2F(\phi_{0},\beta_{0},C_{n},D_{n};\xi,z) \end{eqnarray*}

where $F$ is a solution to the system
\begin{eqnarray*} {1\over
\xi }\partial_{\xi}F = {1\over 2}((\partial_{\xi}\phi)^2 +
(\partial_{z}\phi)^2) , \\ {1\over \xi}\partial_{z}F =
\partial_{\xi}\phi \partial_{z}\phi \, .\end{eqnarray*}

For more details see \cite{CCM}.

In the limit $\xi \to 0$ the minimally coupled scalar field and
the auxiliary field behave as

\begin{eqnarray*} \phi \sim \phi_{*}(z)+ \beta(z)\ln(\xi) , \\ {\tilde \phi}
\sim {\tilde \phi_{*}}(z) + \alpha(z)\ln(\xi) \end{eqnarray*}

where

\begin{eqnarray*} \alpha(z)= \alpha_{0} + {2\over
\pi}\sum_{n=1}^{\infty}B_{n}\cos(n(z-z_{n})) ,\\ \beta(z)=
\beta_{0} + {2\over
\pi}\sum_{n=1}^{\infty}D_{n}\cos(n(z-z_{n}))\end{eqnarray*}

and

\begin{eqnarray*} \phi_{*}(z)= \phi_{0} +  \\
\sum_{n=1}^{\infty}\left(C_{n} + {2\over \pi}D_{n}(\gamma +
\ln({n\over 2})) \right)\cos(n(z-z_{n})) ,\end{eqnarray*}
\begin{eqnarray*}
{\tilde \phi_{*}}(z)= {\tilde \phi_{0}} + \nonumber \\
\sum_{n=1}^{\infty}\left(A_{n} + {2\over \pi}B_{n}(\gamma +
\ln({n\over 2})) \right)\cos(n(z-z_{n})) .\end{eqnarray*}

Here $\gamma$ is the Euler constant.

The asymptotic form of the Charach metric can be written as

$$ ds^2 = A_{s}^2(\xi,z)(dz^2 - d\xi^2) + B_{S}^2(\xi,z)dx^2 +
C_{S}^2(\xi,z)dy^2 $$

where

\begin{eqnarray*} A_{S}(\xi,z) =  \\ e^{\gamma(z)}\xi^{{1\over
4}\alpha^2(z) + \beta^2(z)}\left(e^{{1\over 2}{\tilde
\phi_{*}}(z)}\xi^{{1\over 2}\alpha(z)}  +  e^{-{1\over 2}{\tilde
\phi_{*}}(z)}\xi^{-{1\over 2}\alpha(z)}  \right) ,\end{eqnarray*}

$$ B_{S}(\xi,z)= \xi\left( e^{{1\over 2}{\tilde
\phi_{*}}(z)}\xi^{{1\over 2}\alpha(z)}  +  e^{-{1\over 2}{\tilde
\phi_{*}}(z)}\xi^{-{1\over 2}\alpha(z)} \right) ,$$

$$ C_{S}(\xi,z) = \left( e^{{1\over 2}{\tilde
\phi_{*}}(z)}\xi^{{1\over 2}\alpha(z)}  +  e^{-{1\over 2}{\tilde
\phi_{*}}(z)}\xi^{-{1\over 2}\alpha(z)} \right)^{-1}. $$

The asymptotic behaviour of the scalar fields in the   Charach
solution in the high frequency regime ($n\xi \gg 1$) is

\begin{eqnarray*} \phi \sim \phi_{0} + \alpha_{0}\ln{\xi} + \xi^{-{1\over
2}}H(\xi,z) , \\ {\tilde \phi} \sim {\tilde \phi}_{0} +
\beta_{0}\ln{\xi} + \xi^{-{1\over 2}}{\tilde H}(\xi,z)
\end{eqnarray*}

where

$$ {\tilde H}(\xi,z)\!=\!\! Re\!\!\!\!\! \sum_{n=-\infty,\ne
0}^{\infty}\!\!\! (A_{\mid n\mid} - i B_{\mid n\mid}){e^{-i(z_{n}
- {\pi\over 4})} \over \sqrt{2\pi\mid n\mid}}e^{i(\mid n\mid \xi +
nz)} , $$

$$ H(\xi,z)\! = \!\!Re\!\!\!\!\! \sum_{n=-\infty,\ne
0}^{\infty}\!\!\!\! (C_{\mid n\mid} - i D_{\mid
n\mid}){e^{-i(z_{n} - {\pi\over 4})} \over \sqrt{2\pi\mid
n\mid}}e^{i(\mid n\mid \xi + nz)}. $$

The functions $H(\xi,z)$,  ${\tilde H(\xi,z)}$ satisfy the
D'Alembert equations

\begin{eqnarray*}
\partial^2_{\xi}H(\xi,z) - \partial^2_{z}H(\xi,z) = 0 ,
\\
\partial^2_{\xi}{\tilde H(\xi,z)} - \partial^2_{z}{\tilde H(\xi,z)} =
0 .
\end{eqnarray*}

When $\alpha_{0}=0$ the asymptotic form of the metric is

$$ g_{\mu\nu} = \eta_{\mu\nu} + h_{\mu\nu} $$

where

$$ \eta = diag\left(-4\xi^{2\beta^2_{0}}e^{K\xi}, 4\xi^{2},
{1\over 4}, 4\xi^{2\beta^2_{0}}e^{K\xi} \right) $$

and

$$ h = diag\left(0, 4\xi^{2}H(\xi,z)/4\xi, -{1\over
4}H^{2}(\xi,z)/4\xi \right) .$$

The constant $K$ is given by

$$ K = {1\over 2\pi}\sum^{\infty}_{n=1}\left(A^2_{n} + B^2_{n} +
4(C^2_{n} + D^2_ {n}) \right) .$$

In the case when $\alpha_{0}\ne 0$ the asymptotic form of the
metric is

\begin{eqnarray*} \eta = diag \left(- \xi^{2\beta^2_{0} + {1\over 2}\mid
\alpha_{0}\mid + {1\over 2}\mid\alpha_{0}\mid^2}e^{K\xi},
\xi^{\mid\alpha_{0}\mid + 2}, \right. \\ \left.
\xi^{-\mid\alpha_{0}\mid}, \xi^{2\beta^2_{0} + {1\over 2}\mid
\alpha_{0}\mid + {1\over 2}\mid\alpha_{0}\mid^2}e^{K\xi} \right)
\end{eqnarray*}

and

\begin{eqnarray*} h = diag \left(0, \pm \xi^{\mid\alpha_{0}\mid +
2}H(\xi,z)/\sqrt{\xi}, \right. \\ \left.\mp
\xi^{-\mid\alpha_{0}\mid}H(\xi,z)/\sqrt{\xi}, 0 \right) .
\end{eqnarray*}

\end{appendix}


\begin{thebibliography}{}

\bibitem{LWC} J.Lidsey, D. Wands, E. J. Copeland
                 Phys. Rep., {\bf Vol. 337}, p.343 (2000)


\bibitem{V1} G. Veniziano, Phys. Lett. {\bf B 265}, 287 (1991);
              M. Gasperini,G Veneziano, Astropart. Phys. {\bf 1}, 317 (1993)

\bibitem{BK} J. Barrow, K. Kunze, Phys. Rev {\bf D 56}, 741 (1997)

\bibitem{FLVM} A. Feinstein, R. Lazkoz, M. Vazquez-Mozo, Phys. Rev
                {\bf D 56}, 5166 (1997)

\bibitem{CFLT} D. Clancy, A. Feinstein, J. Lidsey, R. Tavakol,
Phys. Rev. {\bf D60}, 043503-1 (1999)

\bibitem{L}  R. Lazkoz, Phys. Rev
                {\bf D 60}, 104008 (1999)


\bibitem{Gowdy} R. Gowdy, Phys. Rev. Lett. {\bf 27}, 827 (1971);
                 Ann. Phys.(N.Y.){\bf 83}, 203 (1974)

\bibitem{Berger} B. Berger,  Ann. Phys.(N.Y.){\bf 83}, 458 (1974);
                    Phys. Rev. {\bf D 11}, 2770  (1975)

\bibitem{Misner} Ch. Misner, Phys. Rev. {\bf D 8}, 3271  (1973)

\bibitem {WIM} J. Wainwright, W. Ince, J. Marshman, Gen. Rel.
Grav.  Vol.{\bf 10},259 (1979)


\bibitem{C} Ch. Charach,  Phys. Rev. {\bf D 19}, 3516  (1979)

\bibitem{CM}Ch. Charach, S. Malin , Rev. {\bf D 21}, 3284  (1980)

\bibitem{CCM} M. Carmeli, Ch. Charach, S. Malin, Phys. Rep. {\bf
76} (1981)

\bibitem{Giovannini} M. Giovannini,  Phys. Rev. {\bf D 59}, 123518-1
(1999)

\bibitem{BKL} V. Belinskii, I. Khalatnikov, E. Lifhsitz, Adv. Phys.
              {\bf 31}, 639 (1972)


\bibitem{Misner1}Ch. Misner , Phys. Rev. Lett. {\bf 29} 1071 (1969)


\bibitem{ELS}D. Eardley, E. Liang, R. Sachs, J. Math. Phys. {\bf
13}, 99 (1972); J. Isenberg, V. Moncrief, Ann. Phys. (NY) {\bf
199}, 84 (1990)

\bibitem {BKh} V. Belinskii, I. Khalatnikov, Sov. Phys. JETP {\bf
32}, 169 (1971)

\bibitem {Berger1} B. Berger, Phys. Rev. {D 61}, 023508-1 (1999)

\bibitem{NTM} M. Narita, T. Torii, K. Maeda, E-print gr-qc/0003013

\bibitem{Geroch} R. Geroch, J. Math. Phys. {\bf 13}, 394 (1972)

\bibitem{Yazadjiev} S. Yazadjiev, Int. J. Mod. Phys. {\bf D8}, 635
(1999)

\bibitem{BZ} V. Belinskii, V. Zakharov, JEPT {\bf 48}, 985 (1978);
JEPT {\bf 50}, 1 (1980)

\end{thebibliography}
\end{document}